# Stokes' Cradle: Newton's Cradle with Liquid Coating

Donahue, C.M., Hrenya, C.M.*, and Davis, R.H.

*Department of Chemical and Biological Engineering, University of Colorado, Boulder, CO 80309–0424, USA*

Flows involving liquid-coated grains are ubiquitous in nature (pollen capture, avalanches) and industry (air filtration, smoke-particle agglomeration, pharmaceutical mixing). In this work, three-body collisions between liquid-coated spheres are investigated experimentally using a "Stokes' cradle", which resembles the popular desktop toy known as the Newton's cradle. Surprisingly, previous work indicates that every possible outcome was observed in the wetted system except the traditional Newton's cradle (NC) outcome. Here, we are able to experimentally achieve NC via guidance from a first-principles model, which revealed that controlling the volume of the liquid bridge connecting the two target particles is the key parameter in attaining the NC outcome. By independently decreasing the volume of the liquid bridge, we not only achieved NC but also uncovered several new findings. For example, in contrast to previous work on two-body collisions, three-body experiments provide direct evidence that the fluid resistance upon rebound cannot be completely neglected due to presumed cavitation; this resistance also plays a role in two-body systems yet cannot be isolated experimentally in such systems. The herein micro-level description provides an essential foundation for macro-level descriptions of wetted granular flows.

Newton's cradle has long been a popular desktop toy. The outcome is well-known: when a solid sphere at the end of a line of dry, suspended spheres is pulled up the arc and released, it falls and strikes the adjacent sphere, causing the sphere on the opposite end to be ejected from the group with a velocity comparable to the impact velocity of the striker. The toy serves as a micro-level probe into the collisional mechanics that feed into macro-level phenomena of granular flows [1-4]. Inspired by this setup, we have employed what we refer to as the Stokes' cradle to study the mechanics of collisions between spheres coated with a thin layer of viscous liquid, where the Stokes flow of the liquid between the spheres dictates the collisional outcome.

In previous low-Reynolds-number (Stokes flow) work on *immersed* collisions, two-body experiments and corresponding theory revealed that the key dimensionless number is the Stokes number ($St$), which is a ratio of inertial to viscous forces. For a deformable sphere, a coupling of solid mechanics and lubrication forces is able to predict the stick or rebound of a collision, depending on the value of $St$ [5]. Experimental findings of immersed collisions agree well with this *elastohydrodynamic* theory [6-8]. The immersed, two-body collisions have provided the foundation for the studies of *wetted* collisions with a few modifications. Previous works indicated that the pressure upon rebound is significantly below the vapor pressure, leading to assumed cavitation, and thus resistance upon rebound has been neglected [9, 10]. However, it is difficult to experimentally isolate the role of cavitation in two-body systems, since a change in the resistance upon rebound (e.g., via a change in viscosity or thickness of the liquid layer) will also result in a change in the approach resistance. Most experimental configurations have consisted of collisions between a sphere and a wet wall [10-14].

Unlike the Newton's cradle toy, which typically has five dry spheres in a row, our focus is on a wetted, three-sphere system as illustrated in Fig. 1. Consequently, compared to two-body collisions, where the possible outcomes only include stick or bounce, four outcomes are possible in a three-sphere collision. In addition to the Newton's cradle (NC) outcome (where the sphere opposite to the striker sphere separates from the remaining agglomerate), the other possibilities include fully agglomerated (FA, where all spheres stick together), "reverse Newton's cradle" (RNC, where the striker sphere separates but the other spheres stay agglomerated), and fully separated (FS). Thus, both agglomeration and de-agglomeration may be studied. Fig. 2 shows representative experimental snapshots of the spheres after the collision for 12.0 Pa·s oil, chrome-steel spheres, and dripping for 60 s before collisions ("thick" liquid

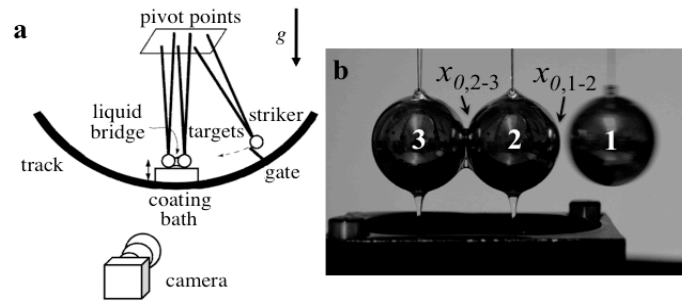

FIG. 1 (a) Stokes' cradle schematic and (b) snapshot of the spheres prior to impact. Three, V-shaped pendulum lines hold 2.54 cm diameter steel spheres. The coating bath is lifted to dip the spheres and then lowered, allowing the spheres to drip as an agglomerated pair with a connecting liquid bridge. The drip time determines the thickness of the oil coating. The dry striker impacts a wet target sphere; thus, a liquid layer is present between each colliding pair of spheres. The striker is released along the arc at different positions, allowing for a range of impact velocities. The impact and post-collisional velocities are measured by a camera. The relevant thicknesses used in model predictions include, $x_{0,1-2}$ and $x_{0,2-3}$, which represent the thickness of the initial liquid layers separating sphere pairs 1-2 and 2-3, respectively, and $x_{f,2-3}$ (not shown), which represents the final liquid thickness for the 2-3 pair, as calculated from the volume of the liquid bridge connecting the targets. The parameters varied include: dry restitution coefficient (0.9 for stainless steel and 0.99 for chrome steel), oil viscosity (12.0 Pa·s and 5.12 Pa·s), oil thickness (60 s [thick layer] and 120 s [thin layer] drip times), and impact velocity (0.1 – 2 m/s).

layer). Here, all parameters are kept constant between the subfigures, except for impact velocity, in which the arrow size represents the relative magnitude. The outcomes as velocity is increased are FA, RNC, and FS, and do not include the NC outcome, which is counter-intuitive given our experience with the toy. In our companion work [15], we presented a new, first-principles model that is able to predict the correct progression of outcomes shown in Fig. 2, including the absence of NC, which is consistent over the large parameter space examined in that work. To date, no other group has published findings on the collisional dynamics between more than two wetted bodies.

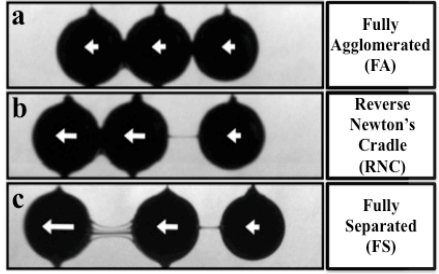

FIG. 2 Snapsnots after a collision using 12.0 Pa·s oil, chrome-steel spheres and 60 s drip time ($x_{0,1-2}$= 410 μm, $x_{0,2-3}$= 14 μm). (a)FA is observed at low velocities, (b)RNC at moderate velocities, and (c)FS at high velocities. Unexpectedly, NC was not initially observed over the wide range of parameters varied.

In the current effort, the previously elusive NC outcome is attained by using the aforementioned model to generate an array of regime maps that identify where in the parameter space the NC outcome is expected. Furthermore, we are now able to show that the outbound resistance plays a critical role in the collisional outcome. A modified experimental method employed here has allowed us to independently change $x_{f,2-3}$ by adjusting the liquid-bridge volume while leaving $x_{0,1-2}$ and $x_{0,2-3}$ fixed (note that $x_{f,1-2} = x_{0,1-2}$, since there is no liquid bridge for the 1-2 pair). This isolation of outbound-resistance effects, which cannot be accomplished in two-body experiments, is detailed below. Moreover, new, counter-intuitive experimental results emerge in this effort, such as producing "stickier" collisions with a thinner liquid layer. The model again provides the physical insight to explain these behaviors. Hence, the following offers a model overview, followed by experimental results that have led to several findings that did not manifest in previous two- or three-body collisions.

To better understand the absence of the NC outcome in our initial series of experiments, we compare our experimental collisions with the theoretical model. The three-sphere collision present in the experiments is approximated as a series of two-sphere collisions in the model. The striker sphere (sphere 1 in Fig. 2) collides with the first target sphere (sphere 2), which subsequently collides with the last target sphere (sphere 3). At this point, the striker sphere may "catch up" and collide again with the first target sphere, which may then collide again with the second target sphere, etc.; if so, the subsequent collisions are also included in the analysis. However, the current work does not include later collisions after the target spheres reach the end of their arcs and reverse direction due to gravity. This two-body interaction sequence is pursued because our work is focused on identifying the dominating physical mechanisms, and preliminary results show only small quantitative differences when a three-body model is employed. Furthermore, and perhaps more importantly, future work will consider dynamic simulations of many-sphere systems. Hard-sphere models, which account for only two spheres colliding at a time, require far less computational power than their soft-sphere counterparts, which incorporate more than two spheres colliding at one time. Nonetheless, the simpler hard-sphere model has been shown to successfully simulate cohesive-particle flows that involve contacts and agglomerates of more than two spheres [16, 17].

The model used here extends the previous models [5, 9] of wetted collisions between only *two* spheres, with two important distinctions detailed later. An analysis of the appropriate dimensionless numbers indicates that Stokes flow prevails (low Reynolds number, $Re = \rho|V|x/\mu < 0.06$) and that capillary forces may be neglected (high capillary number, $Ca = 3\mu a|V|/\sigma x > 3400$). Here, $\rho$ is the fluid density, $V$ is the relative velocity of the two spheres, $x$ is the separation distance between the surfaces of the two spheres, $\mu$ is the fluid viscosity, $a = a_1 a_2/(a_1+a_2)$ is the reduced radius of the spheres, and $\sigma$ is the fluid surface tension. Additionally, air resistance is neglected, as is gravity except to provide the initial velocity of the striker sphere (the long pendulum lines are restricted to small departures from vertical during the collision). The relevant equations of motion for two wetted spheres are provided in our previous work [15] and contain no fitting parameters. As the spheres approach each other, they (i) experience resistance starting at the separation $x_0$ during approach due to lubrication, (ii) may reach a minimum separation and reverse direction due to one of three criteria, and (iii) experience resistance upon rebound until the separation reaches $x_f$, noting that there is not sufficient fluid from the initial layer (1-2 pair) or bridge (2-3 pair) to fill larger gaps. Agglomeration will occur if the initial momentum is not great enough to overcome the total resistance provided by lubrication. On the other hand, rebound past $x_f$ may occur if the initial momentum is large enough that a portion of the kinetic energy becomes stored as elastic deformation rather than lost to viscous dissipation. However, rebound of sphere 1 from sphere 2 occurs more easily than sphere 2 from sphere 3, because of the additional resistance from the excess fluid associated with the liquid bridge (visible in Fig. 1b) between spheres 2 and 3, leading to a bias for the RNC versus NC outcome. The rebound criteria include surface roughness (a measurable quantity corresponding to the dry sphere surface), the elasticity length scale (derived previously from the theory of elastohydrodynamics [5, 9]), and the glass-transition length scale (derived by assuming the viscosity of the oil remains at the atmospheric-pressure viscosity until the glass-transition pressure of the oil is achieved). Unlike existing two-sphere models [5, 9], we have included outbound resistance, which was previously neglected due to assumed cavitation. Specifically, this increased resistance will cause the collisions to become stickier, particularly between the target spheres

(spheres 2 and 3), where the outbound resistance is large due to the relatively large volume of liquid contained in the bridge. The model is able to successfully reproduce the same progression of outcomes as observed in the experiments, as shown in Fig. 2, as well as other observed experimental trends. Further details are given elsewhere[15].

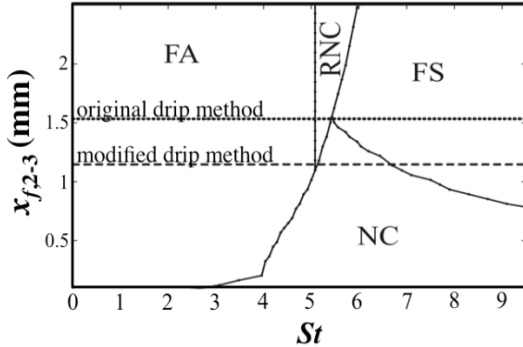

FIG. 3 Model-based regime map of $x_{f,2-3}$ versus $St$. Parameters used here are 12.0 Pa·s oil, chrome-steel spheres and a drip time of 60 s ($x_{0,1-2}$ = 410 μm, $x_{0,2-3}$ = 14 μm). The dotted line (top) represents the spheres dripped as an agglomerated pair, while the dashed line (bottom) the spheres dripped separately. Note that $x_{0,2-3} \ll x_{f,2-3}$ since the two target spheres are pulled together by capillary forces prior to the collision.

Encouraged by the correct model prediction of the experimental trends over a wide parameter space, we have extended the parameter space of the model even further to explore the possibility of achieving the NC outcome. A model-based map of $x_{f,2-3}$ versus $St$ is shown in Fig. 3. When the target spheres reach a rebound criterion, reverse direction and separate, the volume of fluid in the connecting bridge fills the gap between the two target spheres as suction pressure draws in fluid. As described previously [15], $x_{f,2-3}$ is the final liquid thickness between the target spheres (2 and 3), after which rebound is assumed to have occurred with no further resistance and is calculated based on the volume of the liquid bridge. $St$ is defined as $mV_0/6\pi\mu a^2$, where $V_0$ is the initial impact velocity of the striker. Notice that Fig. 3 shows the desired NC outcome in the lower-right corner. However, the top dotted line represents the original value of $x_{f,2-3}$ used in the prior experiments (Fig. 2) and does not include NC. The absence of NC is consistent with experimental observations (FA, RNC, and then FS as impact velocity is increased while holding other parameters constant). This map suggests reducing the value of $x_{f,2-3}$ amply leads to a NC outcome, perhaps due to a reduced viscous resistance as the final fluid layer thickness between the two targets is reduced (more discussion provided later).

In an attempt to experimentally achieve the NC outcome, the final thickness of the liquid layer between the initially motionless spheres ($x_{f,2-3}$) was decreased while all other parameters were kept constant. A smaller final thickness was achieved by modifying the dripping process to yield a smaller volume of the liquid bridge. For results presented thus far and contained in our previous work [15], the two target spheres were dipped in the same coating bath and allowed to drip as an agglomerated pair with a liquid bridge connecting them. To reduce the bridge volume and consequently $x_{f,2-3}$, the target spheres in the current work are separated while undergoing the dripping process and are brought together just prior to the collision. In this way, fluid more easily drains from the pair, decreasing the excess fluid in the bridge ($x_{f,2-3}$ reduced) while maintaining the initial liquid-layer thicknesses ($x_{0,1-2}$ and $x_{0,2-3}$).

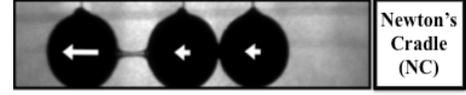

FIG. 4 Snapshot after a collision with a NC outcome using 12.0 Pa·s oil, chrome-steel spheres and 60 s drip time ($x_{0,1-2}$= 410 μm, $x_{0,2-3}$= 18 μm, $x_{f,2-3}$= 1150 μm).

When using this modified dripping method and thus achieving a smaller liquid-layer thickness between spheres 2 and 3 upon rebound (represented in Fig. 3 by the dashed line), the previously missing NC outcome is indeed obtained at intermediate impact velocities, as suggested by the model-based regime map. In particular, outcomes of FA and NC are obtained as $St$ is increased (i.e., going left to right in the regime map of Fig. 3); FS was not observed due to experimental limitations on the maximum velocity (i.e., $St$) that could be achieved. Fig. 4 shows snapshots just before and after the collision for a case where NC was achieved.

The experiments with the reduced value of $x_{f,2-3}$ (using the modified dripping method) show different outcomes that are consistent with model predictions over the range of parameters explored (i.e. impact velocity, oil viscosity, oil thickness, and ball material). Notably, the collisions between stainless-steel spheres, using 12.0 Pa·s oil and the modified (separated spheres) dripping method for 2 minutes, exhibited outcomes of FA at low $St$, RNC and FS at middle $St$, and NC at high $St$, as shown in Fig. 5. On the other hand, NC was never observed for the collisions between spheres using 5.12 Pa·s oil, even when using the modified dripping method and even though this fluid has less viscous resistance, which is also consistent with the model within the experimental uncertainty.

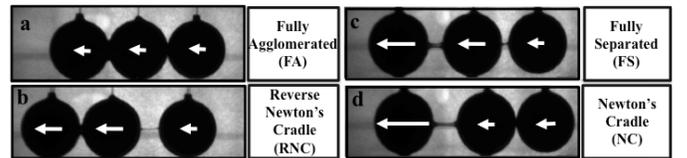

FIG 5 Experimental results of collisions that exhibit all four outcomes. Collisions used 12.0 Pa·s oil, stainless-steel spheres and a drip time of 120 s ($x_{0,1-2}$ = 323 μm, $x_{0,2-3}$ = 11 μm, $x_{f,2-3}$ = 800 μm). (a) FA is observed at low velocities, (b) RNC and (c) FS at moderate velocities, and (d) NC at high velocities, consistent with the model-based regime map (not shown).

Perhaps even more curious than the initial absence of the NC outcome itself is the physical reasoning that eventually leads to its discovery. Consider first the wetted spheres that once displayed FA, RNC and FS as $St$ is increased (Fig. 2). By decreasing the liquid bridge volume and thus $x_{f,2-3}$ (i.e., going "downward" in the regime map of Fig. 3), the resistance between target spheres (2 and 3) decreases. Consequently, it may seem natural that regions of low $St$ that were once FA (2-3 and 1-2 agglomerated) would now separate and exhibit the NC outcome (2-3 separated and 1-2 agglomerated), since no change was made to the liquid layer between the striker and the first target (1-2). However, we found from both experiments and from model predictions, that, as the 2-3 bridge thickness decreases, regions of the regime map (Fig. 3) that were FA remain so (left-hand side), and regions of the regime map that exhibited FS (2-3 separated and 1-2 separated) now exhibit the desired NC configuration (Fig. 4) for the same $St$ (right-hand side). In other words, a change in the resistance between the 2-3 target spheres does nothing to the 2-3 outcome, but rather influences the outcome of the 1-2 pair. The ability of the hard-sphere model to successfully predict the outcome can be traced to the resolution of subsequent binary collisions (when one sphere "catches up" to another after the first series of collisions), and is another testament to the robustness of the model. For example, one way of achieving FS in the model is if sphere 2 rebounds off sphere 1 and sphere 3 rebounds off sphere 2. However, if after this first set of binary collisions, sphere 2 transferred enough momentum to sphere 3 so that sphere 1 has a greater velocity than sphere 2, they will collide again. If they stick together and their velocity is less than sphere 3, a NC outcome will result. Physically, as the striker sphere impacts the targets, the 2-3 liquid bridge dampens the transfer of momentum to sphere 3. Therefore, sphere 2 retains a larger portion of the momentum and does not become agglomerated with sphere 1, which has lost most of its momentum. As the 2-3 bridge thickness decreases, more momentum is transferred to sphere 3, and sphere 2 ends up with less momentum, causing spheres 1 and 2 to agglomerate.

This transition of the FS outcome to NC with decreasing 2-3 bridge thickness at the same impact velocity is also counter-intuitive for another reason. Specifically, in two-sphere collisions, a decrease in the thickness of the liquid layer is associated with more "bounciness" (i.e. a thinner liquid layer results in a transition from agglomeration to rebound at a smaller $St$). A naïve translation to three-sphere collisions may imply that a thinner liquid layer would result in more spheres becoming separated. However, the smaller $x_{f,2-3}$ results in two spheres still agglomerated (NC), while a thicker layer results in all three spheres separated (FS) for the same $St$. So, a thinner liquid layer does not always result in more spheres rebounding, as confirmed by experiments and predictions alike.

In addition to the surprising experimental results already mentioned and explained above via the physical model, the three-sphere collisions examined here provide a rare example of when a more complicated system reveals a physical process that is also important to, but not revealed by, a simpler system. More specifically, previous two-body models predict that the pressure in the liquid gap during rebound drops below the vapor pressure of the oil and thus cavitation was assumed to occur [5, 9]. Therefore, no lubrication resistance upon rebound was included in the previous models and the concept of a final rebound thickness ($x_f$) is thus irrelevant. In two-sphere collisions, the importance of resistance during the rebound phase is difficult to test, since the final thickness cannot be independently changed without also changing the initial thickness (i.e., no liquid bridge as a source of excess fluid exists prior to collision). However, in the more complicated three-sphere collision, the final thickness between the target spheres can be independently changed, since it is controlled via the bridge volume while the initial thickness is controlled by the surface tension that pulls the spheres together. As described above, we found that, when all other parameters were held constant, the decrease in bridge thickness made a qualitative change in the results. Therefore, investigating three spheres instead of two spheres leads to an important conclusion about the basic model for two spheres: outbound resistance plays a major role in the collisional process, even under conditions in which it appears that cavitation may be present.

Understanding the physics of collisions between more than two spheres is necessary not only to understand agglomeration but also de-agglomeration, since de-agglomeration can occur when an existing agglomerate of spheres collide with another body. The plethora of unexpected results described above, such as the initial absence of the NC outcome or more de-agglomeration with a thicker bridge liquid layer, is not possible from two-body studies. Accordingly, the three-body work presented here is an important step as we look toward the macro-behavior of practical, many-body systems.

This work was supported by the National Science Foundation (CBET 0754825) and by the National Aeronautical and Space Administration (NNC05GA48G). C.M.D. would also like to acknowledge fellowship assistance from the Department of Education GAANN Program (P200A060265).